%% file: EE_Cap_EW_UpIntro.tex
\documentclass[final,10pt,twocolumn,a4paper,conference]{IEEEtran}
\IEEEoverridecommandlockouts

\usepackage{amsmath,amsfonts,amssymb,amsbsy,url,verbatim}
\usepackage{times}
\usepackage[english]{babel}
\usepackage[pdftex]{graphicx}
 \usepackage{epstopdf}
  \usepackage{subfigure}
\usepackage{bm}
\usepackage{comment}
\usepackage{cite}
\usepackage{dsfont} 
\usepackage{wrapfig}
\usepackage{algorithm}
\usepackage{algcompatible}
\usepackage[top=1in, bottom=1in, left=0.7in, right=0.7in]{geometry}

\makeatletter
\newcommand\remembertext[2]{
  \immediate\write\@auxout{\unexpanded{\global\long\@namedef{mytext@#1}{#2}}}%
  #2%
}
\newcommand\recalltext[1]{%
  \ifcsname mytext@#1\endcsname
    \@nameuse{mytext@#1}%
  \else
    ``??''
  \fi
}
\makeatother

\newcommand{\diag}{\mathrm{diag}}

\graphicspath{%
     {./figures/}%
}

\begin{document}

\renewcommand{\tablename}{\textsc{TABLE}}

\title{Millimeter Wave Receiver Comparison Under Energy vs Spectral Efficiency Trade-off }
\vspace{-1mm}

\author{
    \IEEEauthorblockN{Waqas~bin~Abbas\IEEEauthorrefmark{1}\IEEEauthorrefmark{2}, Felipe~Gomez-Cuba\IEEEauthorrefmark{1}\IEEEauthorrefmark{3}, Michele Zorzi\IEEEauthorrefmark{1}
    \IEEEauthorblockA{
    \IEEEauthorrefmark{1}DEI, University of Padua, Italy.\\
   \IEEEauthorrefmark{2}National University of Computer and Emerging Sciences, Pakistan.\\
    \IEEEauthorrefmark{3}Stanford University, USA.\\
    \texttt{waqas.abbas@nu.edu.pk,\ gmzcuba@stanford.edu,\ zorzi@dei.unipd.it}
    }
    \vspace{-1.11cm}}
 \thanks{Michele Zorzi was partially supported by NYU-Wireless. This project has received funding from the European Union's Horizon 2020 research and innovation programme under the Marie Sk\l{}odowska-Curie grant agreement No 704837.} 
}

\maketitle

\begin{abstract}
Receivers for mmWave systems suffer from high power consumption in Analog to Digital Converters (ADC), and there is a need to compare the three major receiver architectures: Analog, Hybrid and Digital Combining (AC, HC and DC). Moreover, the specific power consumption figure of merit of ADCs varies significantly between different component designs in the literature, so that comparisons performed for one ADC model --- no matter how representative of the state of the art --- do not necessarily carry over to other ADC designs with different figures of merit. In this work, we formulate a comparison method between AC, HC and DC that can be easily reproduced with different power consumption parameters and provides all information for receiver architecture selection in a compact chart figure. We also present an interpretation of the receiver selection decision problem as a multi-objective utility optimization to find the best Spectral Efficiency (SE) versus Energy Efficiency (EE) trade-off. We use existing results on the achievable rate of AC, HC and DC systems and an Additive Quantization Noise Model (AQNM) of the ADC capacity degradation. For some example commercial component parameters, we show that the usually held belief that DC requires the highest power is not valid in many cases. Rather, either DC or HC alternatively result in the better SE vs EE trade-off depending strongly on the considered component parameters and on the weight assigned to SE vs EE in the utility maximization.
\end{abstract}


\section{Introduction}
\label{sec:intro}
A 1000x increase in capacity and energy efficiency is one of the key requirements of fifth generation (5G) wireless communications.
Millimeter wave (mmWave) communications, on one hand, are expected to enable greatly increased data rates in future wireless systems \cite{KhanFmmWave,5GWillWork}. On the other hand, although beamforming using large antenna arrays can overcome the high path-loss associated to frequencies at these wavelengths, 
using large antenna arrays with a wide bandwidth may lead to high power consumption, especially in the analog to digital converters (ADC) at the receiver, where energy efficiency may become a critical issue. 

Power consumption of an ADC increases linearly with bandwidth and exponentially with the number of bits, and therefore may easily become the main constraint in mmWave multi-antenna receiver technologies \cite{mmWBF2014}. The receiver architectures to reduce power consumption discussed so far in the literature can be categorized in three families. \emph{Analog Combining (AC)} relies on a single Radio-Frequency (RF) chain and ADC. AC consumes the least power and is an attractive choice whenever more versatile digital processing is not really necessary \cite{AlkhateebMIMOSolMag}.  \emph{Hybrid Combining (HC)}  performs combining in both the analog and the digital domains to reduce the number of RF chains and ADCs while still allowing some spatial multiplexing \cite{AyachHC}. \emph{Digital Combining (DC)} with low-resolution ADCs (for example, 1-4 bits) can reach a good power efficiency but at the cost of an increased quantization error \cite{MoHeath1bit,MassMIMO1bit}.
In this work, we study the spectral efficiency (SE) vs. energy efficiency (EE) trade-off for quantized analog, digital and hybrid receiver architectures.  

 The design of energy efficient multiple antenna receivers has been studied in several works. MmWave receivers with low resolution ADCs are studied in \cite{confiwcmcNossekI06,MurrayAGCQuant,Madhow_1bitADC}. The relationship between the number of ADC bits $b$ and the bandwidth $B$ is studied in \cite{OrhanER15PowerCons}, and MIMO with low resolution is treated in \cite{FanULRateLowADC15,ZhangSELowADC}.
 
 Comparisons among the different architectures can be found in \cite{Ahmed16_EE,MyPCCompEW16}. In \cite{Ahmed16_EE}, low resolution HC beamforming algorithms are designed and the energy efficiency is thoroughly studied in comparison with DC. In \cite{MyPCCompEW16}, low resolution AC, DC and HC are compared with each other and it is shown that low-resolution DC may display a better energy efficiency than HC in some scenarios.

 A significant difficulty in the study of energy efficiency in large array receivers is that ADC power values are rapidly changing. A new ADC design has been recently proposed in \cite{NYU_ADC65fJ} with a power consumption one order of magnitude lower than the ``state of the art representative'' ADCs referenced by receiver analyses such as \cite{OrhanER15PowerCons,Ahmed16_EE,MyPCCompEW16}. Three examples of the rapid change of Walden's figure of merit (the energy consumption per conversion step per Hz) of ADCs \cite{FOM} are given in Table \ref{tab:adcparam}. In the present paper, we use the term High Power ADC (HPADC) to refer to the ADC power consumption values featured in mmWave receiver power consumption analyses such as \cite{OrhanER15PowerCons}. We call Intermediate Power ADC (IPADC) the recently proposed ADC circuit \cite{NYU_ADC65fJ}, designed specifically for mmWave receivers. And we call Low Power ADC (LPADC) a best case scenario for ADC power consumption that can be inferred from the exhaustive survey of hardware designs in \cite{ADCs_97-15}.
  
\begin{table}
 \centering
 \caption{ADC Walden's figure of merit ($c$) in different references}
 \label{tab:adcparam}
 \begin{tabular}{rll}
 Scenario & Value & Generation\\ \hline
 HPADC & $494$ fJ/step/Hz & Receiver design state of the art \cite{OrhanER15PowerCons}\\
 IPADC & $65$ fJ/step/Hz & Recently proposed for mmWave \cite{NYU_ADC65fJ} \\
 LPADC & $5$ fJ/step/Hz & Ideal future projection in survey \cite{ADCs_97-15}\\
 \end{tabular}
\end{table}

 Given this extremely dynamic parameter variation, we claim that the previous literature comparing AC, HC and DC in  \cite{Ahmed16_EE,MyPCCompEW16} is incomplete in the sense that the results are only relevant as far as the ``representative ADC'' they selected remains the state of the art. In this work, we have focused on the design of a SE vs EE trade-off comparison chart method that is easily reproducible for different values, rather than generating results only for one chosen representative ADC model. We will give example numeric results using the values of HPADC and LPADC, and the paper is complemented by a web tool where the reader may generate a chart reproducing our analysis method using any other component parameter values \cite{mmWaveADCwebviewer}. Our contribution extends existing receiver analyses and comparisons in the following ways: 
\begin{itemize}
\item First, we take into account engineers preference between SE and EE. We express this as a preference weight in a multi-objective utility maximization problem where the relative weight of SE vs EE is a free parameter. This gives new insights on the choice between HC and DC, complementing the EE comparison in \cite{Ahmed16_EE}.
\item Second, we consider both Uplink and Downlink scenarios with different numbers of antennas. Our results show that in a Downlink scenario where the receiver has fewer antennas, and using the same HPADC as in \cite{OrhanER15PowerCons}, DC provides the best SE vs EE trade-off among all schemes. In Uplink, although HC gives the best standalone EE, DC comes quite close while allowing better SE, and thus the engineers' preference on the EE-vs-SE trade-off determines the ideal receiver.
\item Third, we show that the SE vs EE trade-off is extremely parameter-dependent by comparing charts with HPADC and LPADC. This means that existing results have limited application to the specific ADC reference chosen and cannot be considered as a universal benchmark, even when the reference is very well selected.
\item Fourth, we guarantee the universal reproducibility of our analysis with different component parameters by providing a web tool to complement the paper \cite{mmWaveADCwebviewer}, where researchers can input different parameters and generate their own SE vs EE chart.
\end{itemize}

Additional results and further discussion can be found in the extended version of this paper in \cite{WaqasEConJour16}.

\section{System Model}
\label{sec:model}
We consider a point-to-point mmWave channel with multiple input multiple output (MIMO) antenna arrays with transmit and receive dimensions $N_t$ and $N_r$, respectively. The signal has bandwidht $B$ and a fast fading frequency-flat impulse response so that inter-symbol interference is not present. Our channel model can extend simply to frequency selective systems with multi-carrier modulations and cyclic prefix. We distinguish three mmWave receiver architectures with different analog or digital MIMO characteristics and ADC quantization, i.e., AC, DC and HC. For all three cases, we consider a fully digital non-quantized architecture at the transmitter. The assumption of digital transmitter is motivated by the fact that ADCs typically cause more power consumption constraints than Digital to Analog Converters. We further assume the availability of
channel state information (CSI) both at the transmitter and at the receiver and we design the MIMO processing accordingly.

\subsection{mmWave Quantized Channel}

The general received signal model is as follows: 
\begin{equation}
\label{eq:channel}
\mathbf{y}_q = (1-\eta)(\mathbf{W}_{BB}^H\mathbf{W}_{RF}^H\mathbf{H}\mathbf{W}_t\mathbf{x} + \mathbf{W}_{BB}^H\mathbf{W}_{RF}^H\mathbf{n}) + \mathbf{W}_{BB}^H\mathbf{n}_q
\end{equation}
where the transmitter starts with a transmitted signal $\mathbf{x}$ which may be scalar or in multiple dimensions and has power constraint $\mathrm{E}\left[|\mathbf{x}|^2\right]=P$. The transmitter projects $\mathbf{x}$ onto the $N_t$ array using a unitary precoding matrix $\mathbf{W}_t$. The signal then passes through an $N_t\times N_r$ random mmWave channel $\mathbf{H}$ with distribution defined in \cite{Akdeniz_mmW_CM} and summarized in Table \ref{tab:mmWHpdf}. After propagation through the channel the receiver array captures the signal $\mathbf{H}\mathbf{W}_t\mathbf{x}$ and the Additive White Gaussian Noise $\mathbf{n}\sim\mathcal{CN}(0,N_0\mathbf{I})$. In AC and HC, the received signal may go through an analog combining stage $\mathbf{W}_{RF}$ where only phase-shifting and addition are permitted (each coefficient of the matrix is a phase change $\{\mathbf{W}_{RF}\}_{i,j}=e^{j\theta_{i,j}}\forall i\in\{1\dots N_t\}\forall j\in\{1\dots N_r\}$). The analog signal in each RF chain is converted at the ADCs and finally goes through digital combining $\mathbf{W}_{BB}$. We model the quantization distortion of the signal using the additive quantization noise model (AQNM) \cite{AQNM07}, in which a noise term $\mathbf{n}_q$ is added before digital combining. We denote by $\eta$ the inverse of the signal-to-quantization noise ratio, which is inversely proportional to the square of the resolution of an ADC.
For a Gaussian input distribution, the values of $\eta$ for $b \leq 5$ (where $b$ is the number of ADC bits) are listed in Table \ref{tab:etavsb}, and for $b > 5$ can be approximated by $\eta = \frac{\pi \sqrt{3}}{2} 2^{-2b}$ \cite{FanULRateLowADC15}.

\begin{table}
     \centering
     \caption{mmWave Channel probability distribution \cite{Akdeniz_mmW_CM}
     $$\mathbf{H} = \sqrt{\dfrac{N_{t}N_{r}}{\rho N_cN_p}}\sum_{k=1}^{N_c}\sum_{\ell=1}^{N_p}g_{k,\ell}\mathbf{a}_{r}(\phi_{k}+\Delta\phi_{k,\ell}) \mathbf{a}_{t}^H(\theta_k+\Delta\theta_{k,\ell})$$}
      \label{tab:mmWHpdf}
     \begin{tabular}{r|r|l}
     Macroscopic Pathloss in dB & $\rho_{LOS}$ & $ 61.5+20\log (d)+\xi$\\
     (Line of Sight at distance $d$) &$\xi$&$\sim\mathcal{N}(0,5.8)$\\\hline
     Macroscopic Pathloss in dB & $\rho_{NLOS}$&$72+29.2\log (d)+\xi$\\
     (Non-LOS at distance $d$)&$\xi$ &$\sim\mathcal{N}(0,8.7)$\\\hline     
     Number of Scattering Clusters & $N_c$ & $\sim\mathrm{Poisson}(1.9)$ \\ 
     Clust. Central Angle of Arrival & $\phi_{k}$ & $\sim\mathrm{U}[0,2\pi]$ \\
     Clust. Ctrl. Ang. of Departure & $\theta_k$ & $\sim\mathrm{U}[0,2\pi]$ \\\hline 
     Number. of Paths per Cluster & $N_p$ & 20 \\
     Small scale scattering per path & $g_{k,\ell}$ & $\mathcal{CN}(0,1)$\\
     Path Differential AoA & $\Delta\phi_{k,\ell}$ & $\sim\mathcal{N}(0,10^o)$ \\
     Path Differential AoD & $\Delta\theta_{k,\ell}$ & $\sim\mathcal{N}(0,10^o)$ \\\hline
     Linear $N$-antenna Array & $\mathbf{a}(\theta)$ & $\left(\begin{array}{c}1\\e^{j\sin(\theta)\frac{1}{N}} \vspace{-.1in}\\
    \vdots\\e^{j\sin(\theta)\frac{N-1}{N}}\end{array}\right)$
     \end{tabular}
 \vspace{-.1in}
 \end{table}
\begin{table}
     \centering
     \caption{$\eta$ for different values of $b$ \cite{FanULRateLowADC15}}\vspace{-3mm}
     \label{tab:etavsb}
     \begin{tabular}{r|lllll}
         $b$ &  1 & 2 & 3 & 4 & 5\\ 
         \hline
         $\eta$  & 0.3634 & 0.1175 & 0.03454 & 0.009497 & 0.002499 \\
     \end{tabular}
 \vspace{-.1in}
 \end{table}

\subsection{Digital Combining (DC) Receiver}
\begin{figure}
 \centering
 \includegraphics[width=0.85\columnwidth]{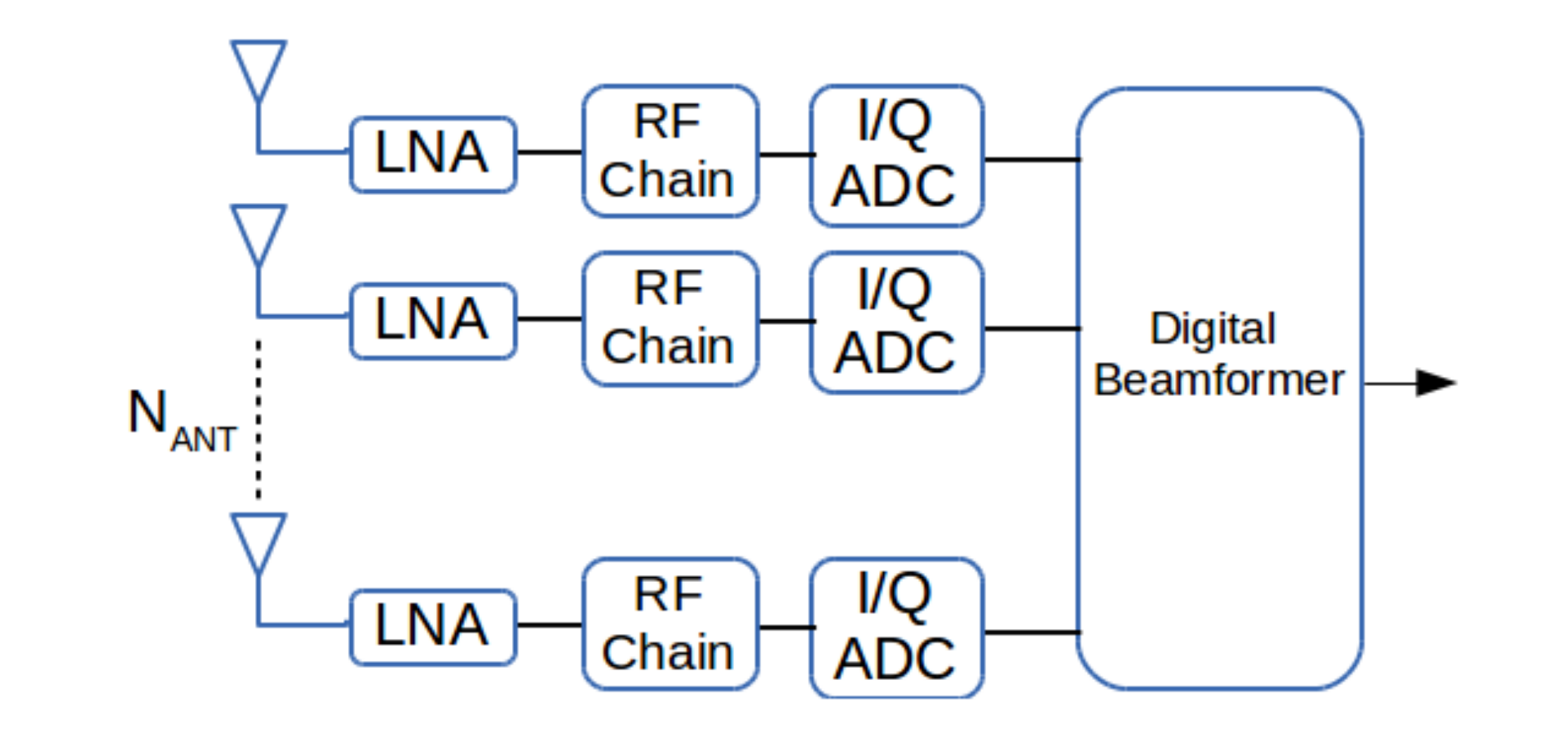}
 \caption{DC Receiver}\vspace{-4mm}
 \label{fig:DC}
\end{figure}

The fully digital receiver is shown in Figure \ref{fig:DC}. DC has a dedicated RF chain per antenna element, and in the channel model \eqref{eq:channel} we have that $\mathbf{W}_{t}$ and $\mathbf{W}_{BB}$ are $N_t\times N_t$ digital precoding and $N_{r}\times N_{r}$ digital combining matrices, whereas analog processing is absent so that $\mathbf{W}_{RF}=\mathbf{I}_{N_r}$.

We compute the DC matrices using the singular value decomposition of the channel matrix $\mathbf{H} = \mathbf{U}\Sigma\mathbf{V}^H$. By applying precoding $\mathbf{W}_t = \mathbf{V}$ and combining $\mathbf{W}_{BB} = \mathbf{U}^H$, we create an equivalent diagonal channel where the transmit power $P$ may be allocated across the singular values $\Sigma$. 
With this formulation, the rate maximization problem with DC is given as
\begin{equation}
\begin{split}
	C_{DC}  =  \mathrm{E}_{\mathbf{H}}\bigg[ \max_{\mathbf{R}_{\mathbf{xx}}} B\log_2\det\bigg|\mathbf{I}+{(1-\eta)\Sigma\mathbf{R}_{\mathbf{xx}}\Sigma^H}\\
	 {({N_o\mathbf{I}} + 
\eta\mathbf{U}^H\diag(\mathbf{U}\Sigma\mathbf{R}_{\mathbf{xx}}\Sigma^H\mathbf{U}^H)\mathbf{U})^{-1}}\bigg|\bigg]	 
	\label{eq:Cq_DC}
	\end{split}
\end{equation}
where the input covariance matrix $\mathbf{R}_{\mathbf{xx}}$ that maximizes the rate is obtained using the water-filling algorithm.

The power consumption of the DC architecture is
\begin{equation}
   P_{Tot}^{DC} =  N_{r}(P_{LNA} + P_{RF} + 2P_{ADC})
   \label{eq:P_DC}
\end{equation}
where $P_{RF} =  P_M + P_{LO} + P_{LPF} + P_{BB_{amp}}$.

In this paper we use an example of component power consumptions detailed in Table \ref{tab:devicepowers}, but other values may also be applied depending on the advances in hardware design. $P_{ADC}$ increases exponentially with $b$ and linearly with $B$ and with the ADC Walden's figure of merit $c$ \cite{ADC_b_B}. The power consumption of all the other components is independent of the bandwidth and the number of bits. 

Finally, we define the EE of each receiver as $\mathrm{EE} \triangleq \dfrac{C}{P_{Tot}}$, where $C$ is the achievable rate and $P_{Tot}$ represents the total power consumption.

\begin{table}
\centering
\caption{Power consumption of each device}
\label{tab:devicepowers}
\begin{tabular}{rll}
Device & Notation & Value\\ \hline
Low Noise Amplifier (LNA)\cite{Rx_Pow_LNA_PS_C} & $P_{LNA}$ & $39$ mW\\
Splitter and Combiner \cite{Rx_Pow_LNA_PS_C}& $P_{SP}$ and $P_{C}$ & $19.5$ mW each\\
Phase shifter \cite{KongPhDPSPow,LPowPS0mW}& $P_{PS}$ & $2$ mW or $0$\\
Mixer \cite{Rx_Pow_60GHz}& $P_{M}$ & $16.8$ mW\\
Local oscillator \cite{RialandHeath} & $P_{LO}$ & $5$ mW\\
Low pass filter \cite{RialandHeath}& $P_{LPF}$ & $14$ mW\\
Base-band amplifier \cite{RialandHeath}& $P_{BB_{amp}}$ & $5$ mW\\
ADC \cite{ADC_b_B} & $P_{ADC}$ & $cB2^{b}
$\\
\end{tabular}
\end{table}

\subsection{Analog Combining (AC) Receiver}
\begin{figure}
 \centering
 \includegraphics[width=0.85\columnwidth]{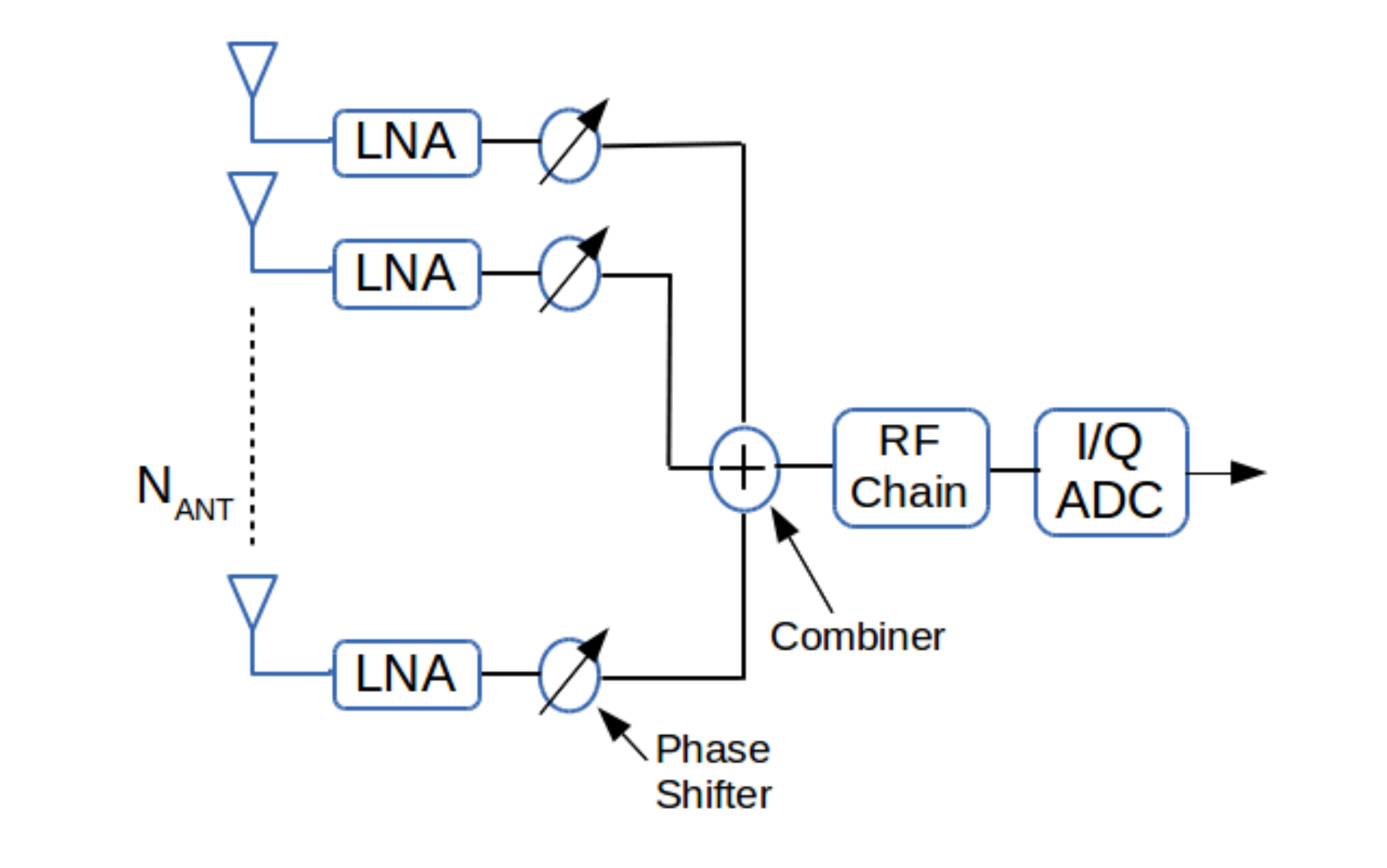}
 \caption{AC Receiver}\vspace{-4mm}
\label{fig:AC}
\end{figure}
The fully analog receiver is shown in Figure \ref{fig:AC}. In contrast to a DC, all receiver processing is performed in the analog domain and only one RF chain and ADC pair are needed. We have that in \eqref{eq:channel}, $\mathbf{W}_{t}$ and $\mathbf{W}_{RF}$ are $1\times N_t$ digital precoding and $N_r\times 1$ analog combining matrices, and digital combining is absent ($\mathbf{W}_{BB}=1$).

The rate maximization problem with AC is given as
\begin{equation}
\begin{aligned}
	C_{AC}  = \mathrm{E}_{\mathbf{H}}\bigg[ \underset{\mathbf{w}_{r},\mathbf{w}_t}  {\max}\ &B\log_2\bigg(1+\dfrac{(1-\eta)|\mathbf{w}_r^H\mathbf{H}\mathbf{w}_t|^2P}{{N_o} + \eta|\mathbf{w}_r^H\mathbf{H}\mathbf{w}_t|^2P}\bigg) \bigg]\\
	 s.t.\ &|w_{r,i}| = \frac{1}{\sqrt{N_r}}, \;||\mathbf{w}_t||^2 = 1,	     
\end{aligned}
	\label{eq:Cq_AC}
\end{equation}
Here we may first work out the digital precoding, which only has a unitary power constraint. We choose $\mathbf{w}_t$ to maximize the gain under any given value of $\mathbf{w}_r$ using $\mathbf{w}_t = \dfrac{\mathbf{H}^H\mathbf{w}_r}{||\mathbf{H}^H\mathbf{w}_r||^2}$. Secondly, in unconstrained circumstances the optimal $\mathbf{w}_r$ would be the maximum left eigenvector ($\mathbf{u}_{max}$) of $\mathbf{H}$, however, due to the constant amplitude constraint $\mathbf{w}_r$ can only be a phase projection of it, i.e.
\begin{equation}
\label{eq:optABFvec}
 \tilde{\mathbf{w}}_r^H=\frac{1}{\sqrt{N_r}}(e^{\measuredangle u_{\max}^1},e^{\measuredangle u_{\max}^2}, \dots e^{\measuredangle u_{\max}^{N_r}})^T
\end{equation}

The power consumption of AC is evaluated as
\begin{equation}
   P_{Tot}^{AC} =  N_{r}(P_{LNA}+P_{PS}) + P_{RF} + P_C+ 2P_{ADC}
   \label{eq:P_AC}
\end{equation}

\subsection{Hybrid Combining (HC) Receiver}
\begin{figure}
 \centering
 \includegraphics[width=0.81\columnwidth]{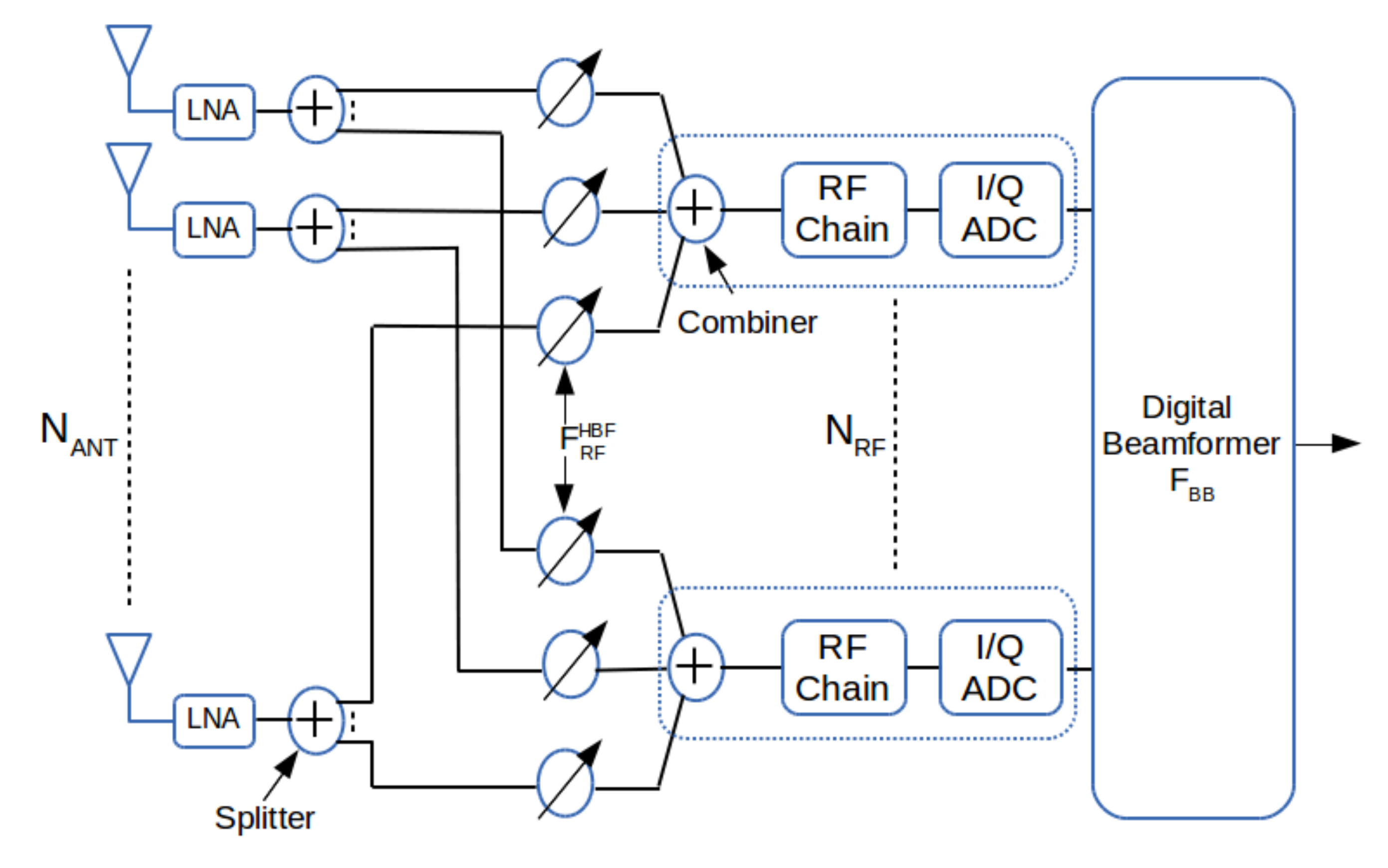}
 \caption{HC Receiver}\vspace{-0mm}
\label{fig:HC}
\end{figure}

The hybrid receiver is shown in Figure \ref{fig:HC}. We have that in \eqref{eq:channel}, $\mathbf{W}_{t}$, $\mathbf{W}_{RF}$ and $\mathbf{W}_{BB}$ are $N_{RF}\times N_t$ digital precoding, $N_r\times N_{RF}$ analog combining and $N_{RF}\times N_{RF}$ digital combining matrices, respectively. The analog combining is obtained using Algorithm \ref{algo:APM} from \cite{TroppAPM05}, and the digital precoding and combining use the Singular Value Decomposition as in the DC model, over an equivalent channel $\mathbf{W}_{RF}\mathbf{H}$ with dimensions $N_t\times N_{RF}$

\begin{algorithm}
\small
\caption{Alternate projection $\mathbf{W}_{RF}$ design}
\label{algo:APM}
\begin{algorithmic}
\STATE {Initialize $\mathbf{W}_{SU}$ = [$\mathbf{u}_1\dots \mathbf{u}_{N_{RF}}$] $\in \mathbf{U}$,where $\mathbf{H} = \mathbf{U}\Sigma\mathbf{V}^H$}
\WHILE { not converging }
  \STATE {[$\tilde{\mathbf{W}}_{RF}]_{ij} = \frac{1}{\sqrt{N_r}}\mathrm{exp}({j\measuredangle[\mathbf{W}_{SU}]_{ij}}$), $\forall i,j$ }
  \STATE {$\mathbf{W}_{SU} = (\tilde{\mathbf{W}}_{RF}\tilde{\mathbf{W}}_{RF}^*)^{-\frac{1}{2}}\tilde{\mathbf{W}}_{RF}$}
\ENDWHILE
\end{algorithmic}
\end{algorithm}
The achievable rate maximization with HC is given as  
\begin{equation}
\begin{split}
	C_{HC}  = \mathrm{E}_{\mathbf{H}}\bigg[\max_{\mathbf{R}_{\mathbf{xx}}} B\log_2\det\bigg|\mathbf{I}+{(1-\eta)\Sigma\mathbf{R}_{\mathbf{xx}}\Sigma^H} \\
	 {({N_o\mathbf{I}} + \eta\mathbf{U}^H\diag(\mathbf{U}\Sigma\mathbf{R}_{\mathbf{xx}}\Sigma^H\mathbf{U}^H)\mathbf{U})^{-1}}\bigg|\bigg]	 
	\label{eq:Cq_HC}
\end{split}
\end{equation}
Note that $C_{HC}$ is upper bounded by $C_{DC}$ due to the fact that $\mathbf{W}_{RF}$ has analog constraints and $N_{RF}\leq N_{r}$.

Finally, the power consumption of HC is evaluated as
\begin{equation}
\begin{split}
   P_{Tot}^{HC} = &\ N_{r}(P_{LNA} + P_{SP} + N_{RF}P_{PS})
    \\&
+ N_{RF}(P_{RF}+P_C + 2P_{ADC} )
\end{split}
\label{eq:P_HC}   
\end{equation} 

\section{SE vs EE Analysis and Examples}
\label{sec:SimAnalysis}
We plot the EE vs SE curves for each receiver as $b$ increases from 1 to 8 at increments of 1. The top of the chart corresponds to highest SE and the rightmost points correspond to highest EE. Thus, an ideal goal would be to design a receiver that is as close to the top right corner of the chart as possible. Nonetheless, this intuitive guideline is not sufficient to capture the needs of receiver design when a trade-off between EE and SE exists. We construct a multi-objective utility optimization interpretation of the receiver selection problem that serves to discuss the interpretation of the chart. Note that this optimization is easy to solve due to the small search space, and its value is not in the result itself but in the interpretation it gives to the chart. 

We consider a free parameter $\alpha\in[0,1]$ that represents the receiver designer's preference between higher EE and higher SE. The ``receiver utility'' according to the designer's preference can be expressed and maximized as 
\begin{equation}
U=\max_{\{HC,DC,AC\}}\max_{b\in\{1\dots8\}} \alpha \mathrm{EE} + (1-\alpha) \mathrm{SE}
\label{eq:UtFun}
\end{equation}
where $\alpha=0$ is SE maximization and $\alpha=1$ is EE maximization. Since the problem is relatively easy, we can obtain the set of solutions for the entire range of parameters $\alpha\in[0,1]$, where the solutions in the set correspond to all the receiver designs that are valuable for some kind of receiver designer preference. Receivers not in the set of  solutions for all $\alpha\in[0,1]$ are receivers that would never be preferred by any designer. 

We obtain SE vs EE charts for AC, HC and DC receivers with $b$ ranging from $1$ to $8$ in four scenarios, i.e., combining Downlink and Uplink with low and high SNR. We highlight the points that are utility maximization solutions in the SE-vs-EE charts to distinguish the most valuable receiver designs from the rest. For Uplink, we use $N_t=16$ and $N_r=64$, and for Downlink, we use $N_t=64$ and $N_r=16$. For high SNR we consider $0$ dB before combining, which is about $100$ m distance in LOS mmWave pathloss, and for low SNR we consider $-20$ dB which is about the same distance with NLOS. We have also introduced different numbers of RF chains in HC to observe the impact of a good selection of the dimension of the hybrid system. Increasing $N_{RF}$ increases SE but also the power consumption, potentially leading to a drop in EE.

We give two examples with two different sets of component parameters\footnote{Although we did not fabricate the receivers (i.e., AC, DC or HC) in hardware, the power consumption model we employ is based on hardware designs considered in recent papers, e.g., \cite{OrhanER15PowerCons,MyPCCompEW16}.}:
\begin{itemize}
 \item The High Power ADC (HPADC) model is based on an existing device that supports sampling at Gs/s and has been referenced in related literature such as \cite{OrhanER15PowerCons}. In order to give HC Phase Shifters appropriate power consumption values, we pair the HPADC model with an existing PS model, i.e., with $P_{PS}=2$ mW, referenced in \cite{KongPhDPSPow}. In this example we focus on comparing the different receivers and on the impact of Downlink/Uplink in high/low SNR scenarios. 
The results validate the observations from \cite{OrhanER15PowerCons,Ahmed16_EE} and also provide some additional insights about HPADC. 
 \item The Low Power ADC (LPADC) model considers a likely best-case scenario deduced from the hardware survey in \cite{ADCs_97-15}. Likewise, we pair this best-case scenario for ADCs with a best case phase shifter model, with negligible power consumption $\sim0$ mW, as in \cite{LPowPS0mW}. For the second example, we focus on displaying how critically parameter-dependent the receiver EE-vs-SE trade-off comparison is. We show how the observations from the same example change drastically and the results in \cite{OrhanER15PowerCons,Ahmed16_EE} for HPADC cannot be generalized to different ADC components in the literature. 
\end{itemize}

Additional results and more examples and discussion can be found in the extended version of this paper \cite{WaqasEConJour16}.

\subsection{Example 1: High power ADC characteristics}

\begin{figure*}[t]
\centering
  \subfigure[High SNR Downlink]{
    \includegraphics[width=.85\columnwidth]{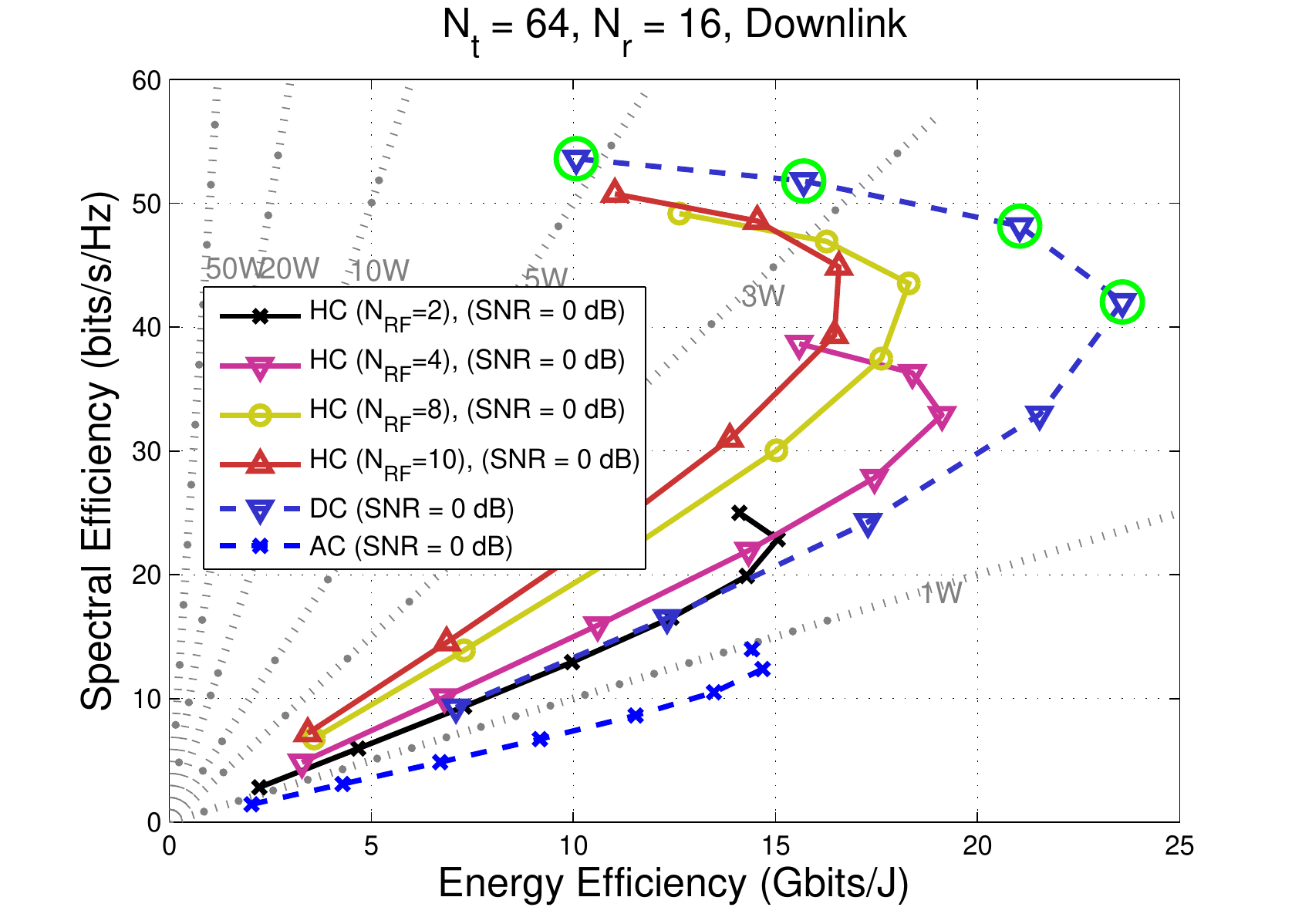}
    \label{fig:DLVarRFHSnr}
    }
    \hspace{.1\columnwidth}
    \subfigure[High SNR Uplink]{
    \includegraphics[width=.85\columnwidth]{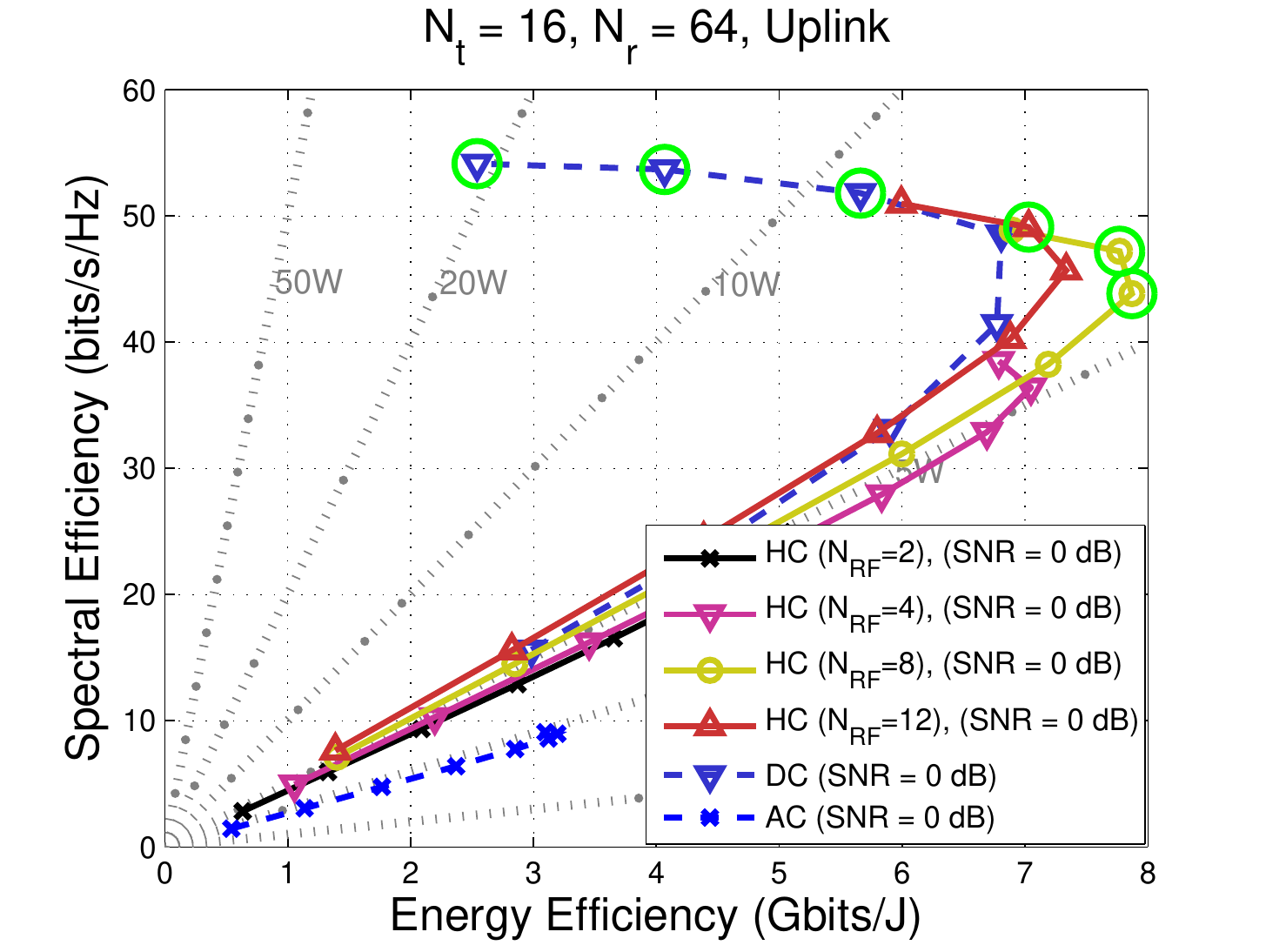}
    \label{fig:ULVarRFHSnr}
    }
      
 \subfigure[Low SNR Downlink]{
    \includegraphics[width=.85\columnwidth]{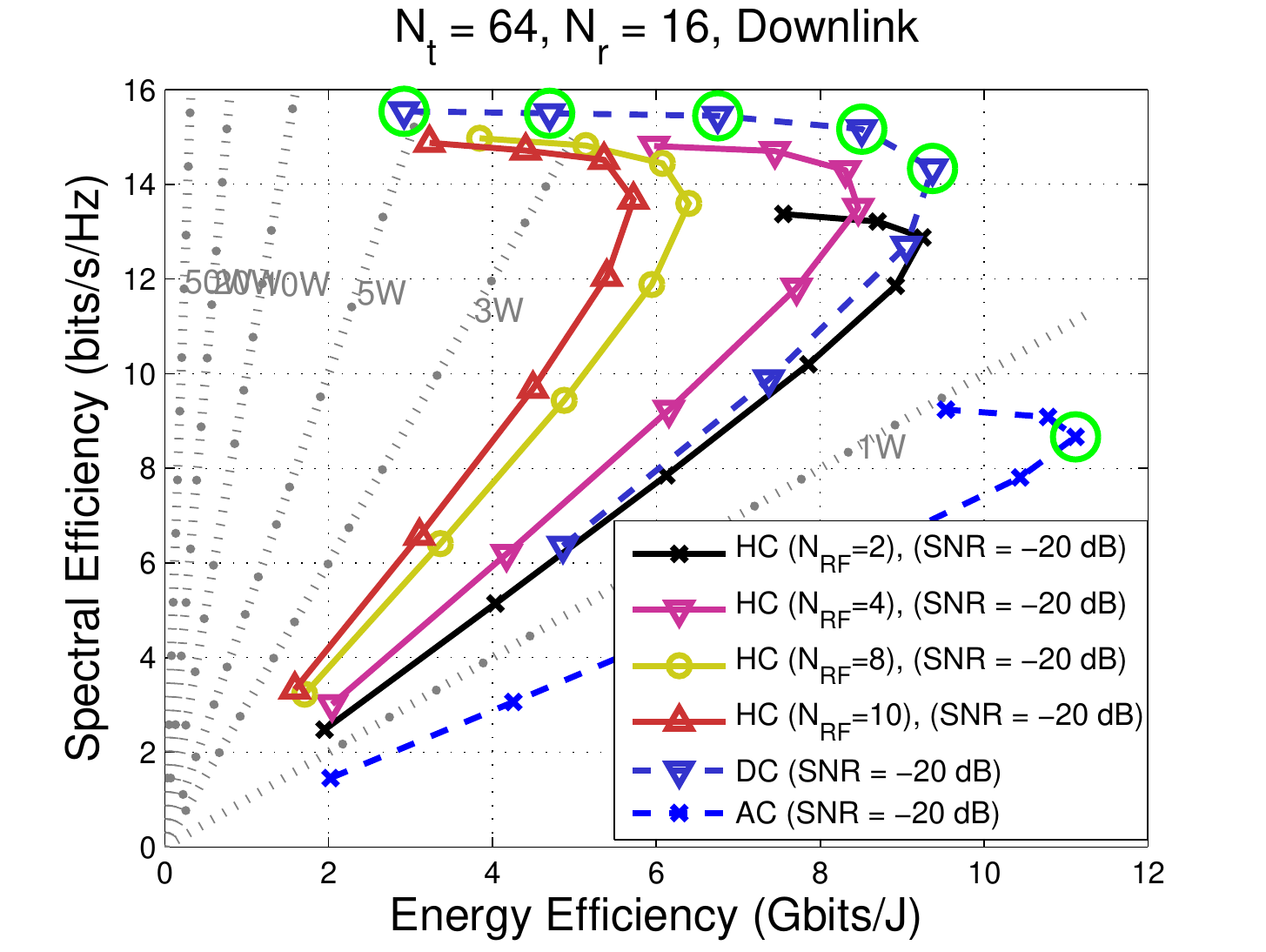}
    \label{fig:DLVarRFLSnr}
    }
    \hspace{.1\columnwidth}
  \subfigure[Low SNR Uplink]{
    \includegraphics[width=.85\columnwidth]{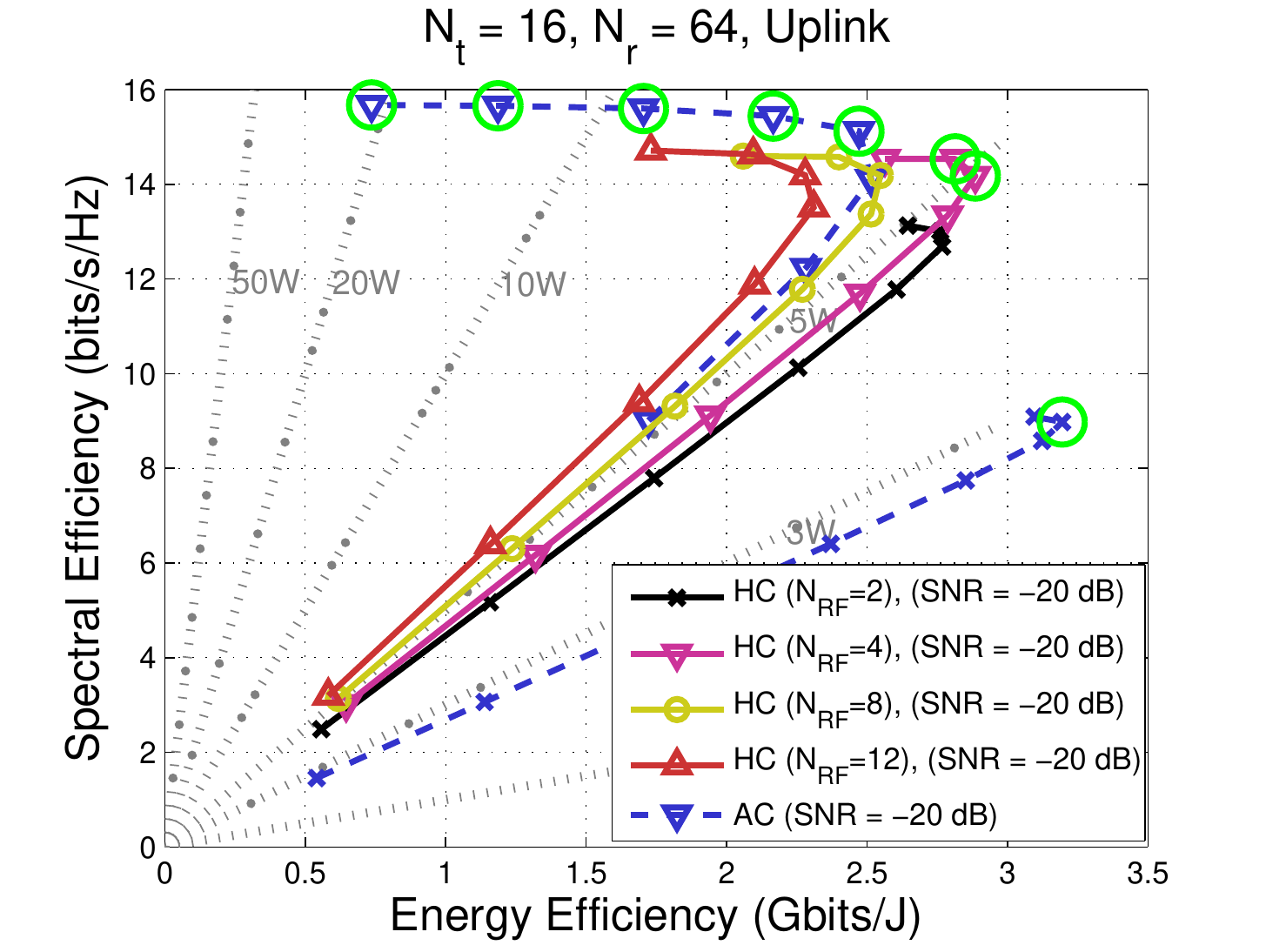}
    \label{fig:ULVarRFLSnr}
    }
  \caption{SE vs. EE comparison for AC, DC and HC schemes with a HPADC model and with $N_{RF} = 2,4,8,10,12$ for HC. In the Downlink, $\max(N_{RF})$ is set to 10 as a further increase would only decrease the EE of HC without any significant SE improvement.}\vspace{-4mm}
    \label{fig:VarRF}
\end{figure*}

We plot the SE vs EE for the three receiver architectures with HPADC in Fig. \ref{fig:VarRF}. Since ADCs have high power consumption, as $b$ is increased the lines first reach upward and right, and then steer left and EE returns to the left corner while SE continues rising. This is consistent with results in \cite{Ahmed16_EE,WaqasEConJour16} that show that SE is monotonic in $b$ whereas EE reaches a maximum and then drops. Note that the points maximizing \eqref{eq:UtFun} for different values of $\alpha\in[0,1]$ are highlighted, so the set of highlighted points in the chart can be regarded as a collection of ``all the interesting receivers''.

Note that the number of bits that maximizes SE is not necessarily the same that maximizes EE. For instance, for DC in Figure \ref{fig:DLVarRFHSnr} the point in the top left corner corresponds to $b = 8$ and maximizes SE (it is highlighted for $\alpha = 0$), whereas the rightmost point correspoinding to $b = 5$ maximizes EE (it is highlighted for $\alpha = 1$). Notice also that the number of bits that maximizes either magnitude varies for different receivers. For example AC, HC with $N_{RF}=4$, and DC achieve maximum EE with $b = 7,6$ and $5$, respectively. 

In Fig. \ref{fig:DLVarRFHSnr} for high SNR Downlink we have a result in which DC completely outperforms all other schemes. DC with 4 or 5 bits has the best EE, and with 8 bits has the best SE. Note that only DC points are selected across all values of $\alpha\in[0,1]$, always achieving the best EE vs SE trade-off. The higher number of antennas in Uplink causes an increase in power consumption that affects DC more severely. In Fig. \ref{fig:ULVarRFHSnr} the high SNR Uplink HC using $N_{RF}=8$ with 6-8 bits achieves greater or equal EE than the best EE of DC. However, HC is only selected with $6-7$ bits for the lowest values of $\alpha$, whereas DC not only is chosen for maximum SE ($b=8$), but also offers the best trade-off with good SE and good EE  for mid-range values of $\alpha$ ($b=5,6$).

To observe the impact of SNR in Downlink we compare Figures \ref{fig:DLVarRFHSnr} and \ref{fig:DLVarRFLSnr}. At high SNR spatial multiplexing offers significant rate gains, and HC SE rises significantly as $N_{RF}$ grows from 2 to 10 while EE does not contract very much. The HC scheme with the best EE uses $N_{RF}=4$ and $b=6$, yet DC still manages to completely outperform HC with any number of RF chains. At low SNR, however, there is no significant spatial multiplexing advantage, and HC SE grows when $N_{RF}$ steps from 2 to 4, but becomes flat thereafter, while EE keeps retracting. Nonetheless, it is at low SNR where we observe a non-DC scheme to get a point in the optimal set for the first time. This scheme is AC which is very well suited for low-SNR scenarios because it pools all the transmitter energy into the strongest propagation direction and has very low component power consumption. Notice that for HC schemes the optimal number of RF chains is SNR-dependent, which is an additional handicap for the use of HC designs, since usually the same hardware design is used in all the devices at different distances in a network. The only Downlink scenario where HC achieves an EE similar to DC is with $N_{RF}=2$ and $b=2$, and yet its SE is $6\%$ lower.

\begin{figure*}[t]
\centering
  \subfigure[High SNR Downlink]{
    \includegraphics[width=.85\columnwidth]{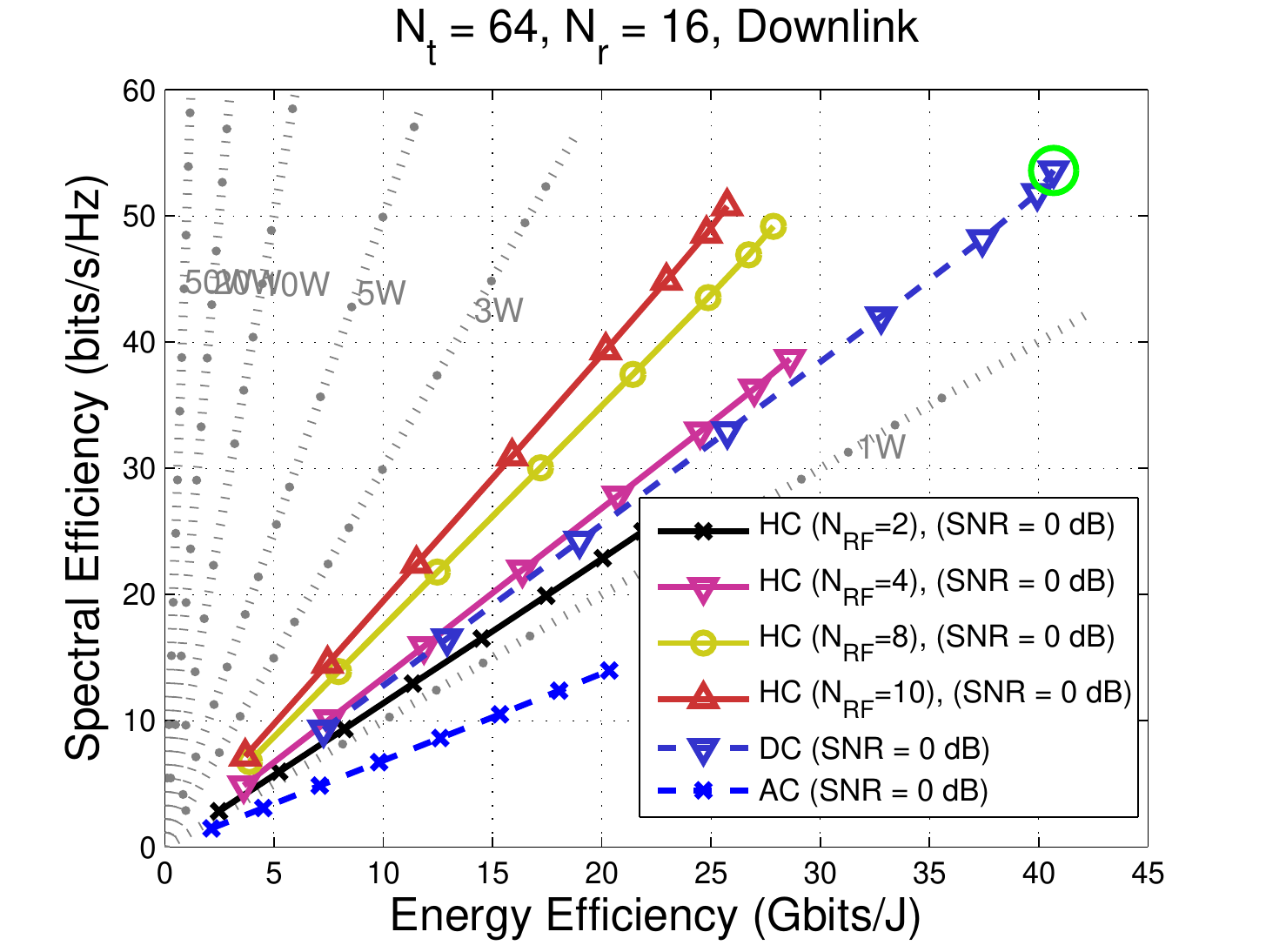}
    \label{fig:DLVarRFHSnrLPADC}
    }
    \hspace{.1\columnwidth}
  \subfigure[High SNR Uplink]{
    \includegraphics[width=.85\columnwidth]{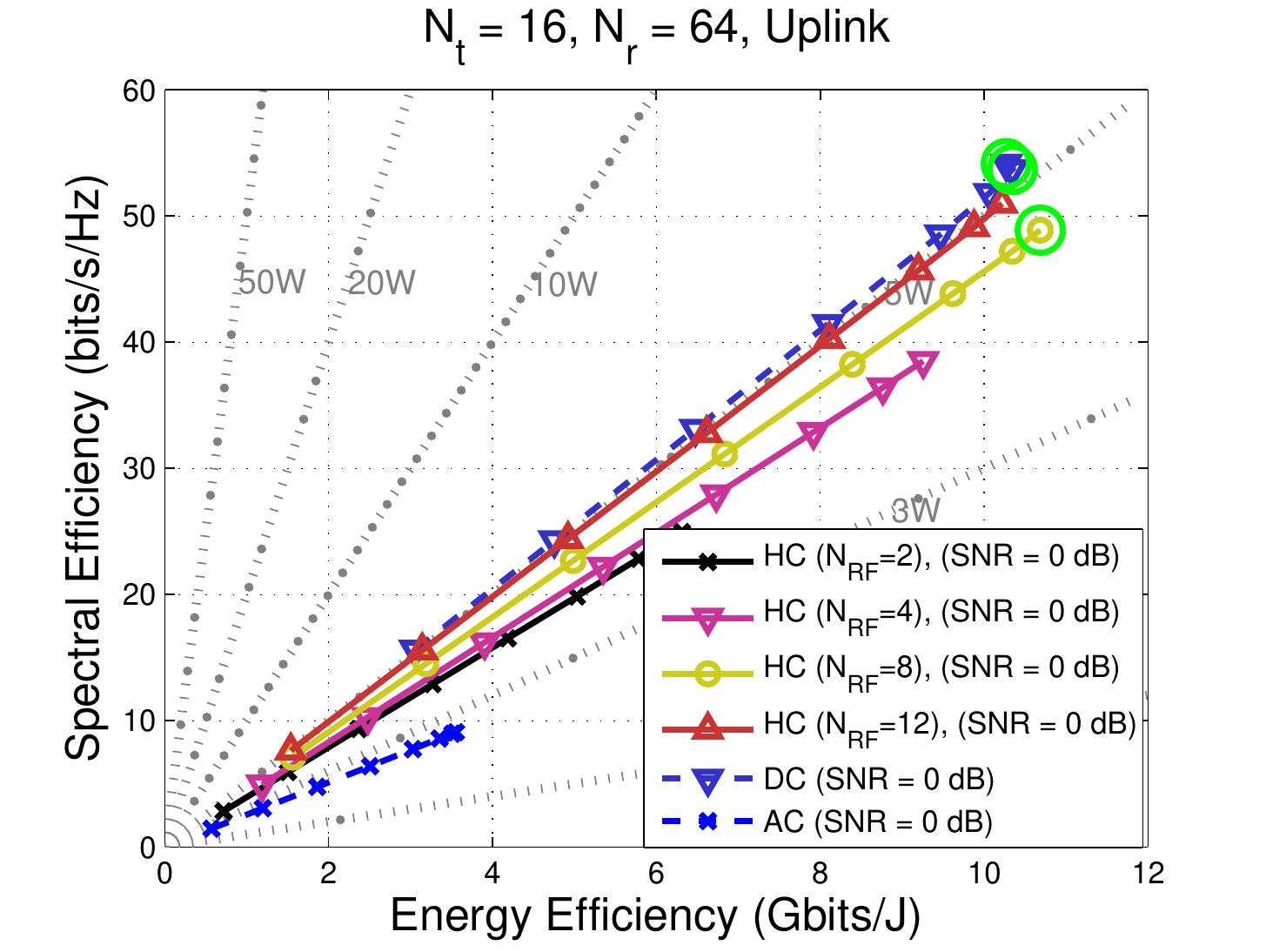}
    \label{fig:ULVarRFHSnrLPADC}
    }
     
  \subfigure[Low SNR Downlink]{
    \includegraphics[width=.85\columnwidth]{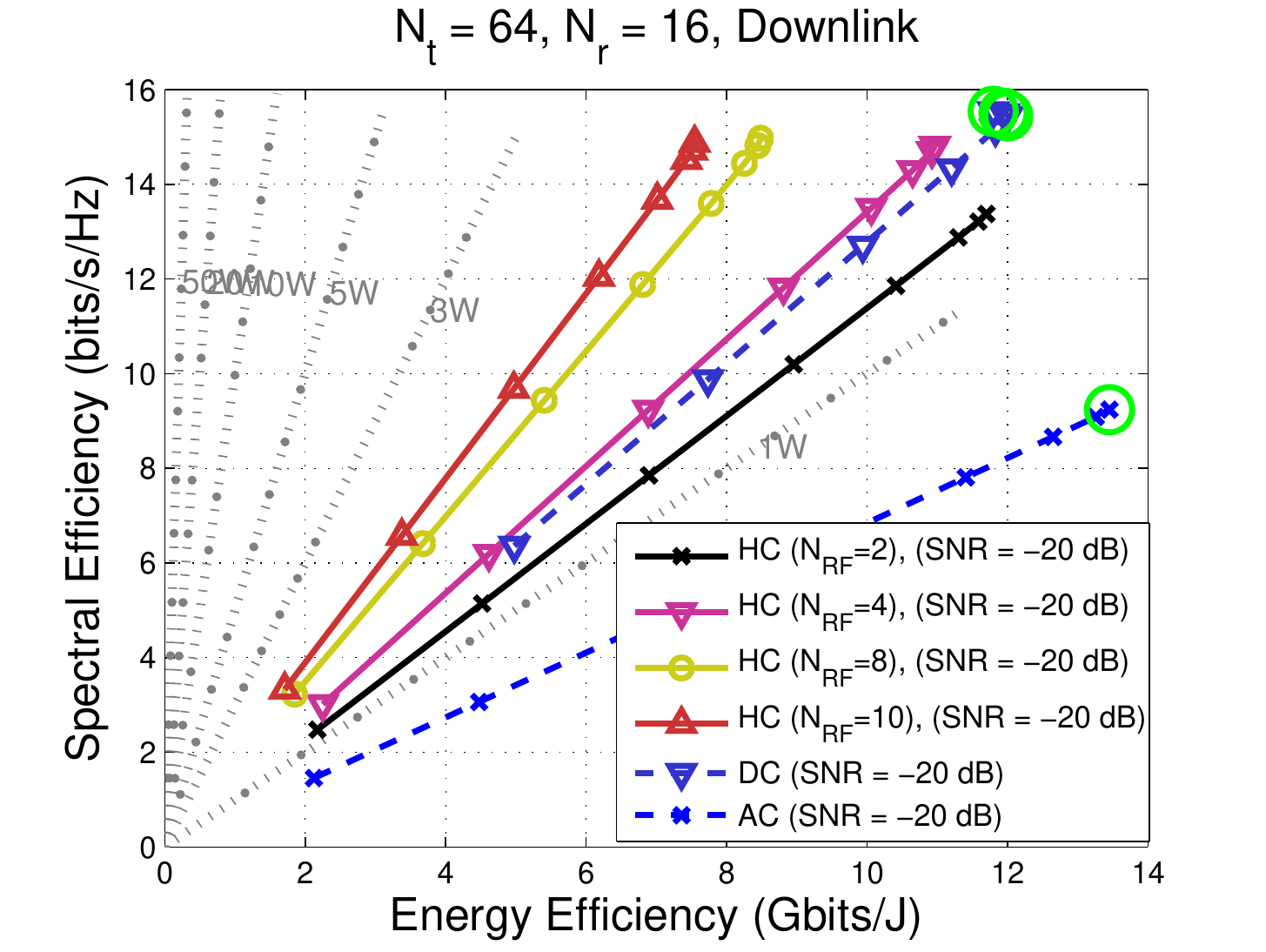}
    \label{fig:DLVarRFLSnrLPADC}
    }
    \hspace{.1\columnwidth}
  \subfigure[Low SNR Uplink]{
    \includegraphics[width=.85\columnwidth]{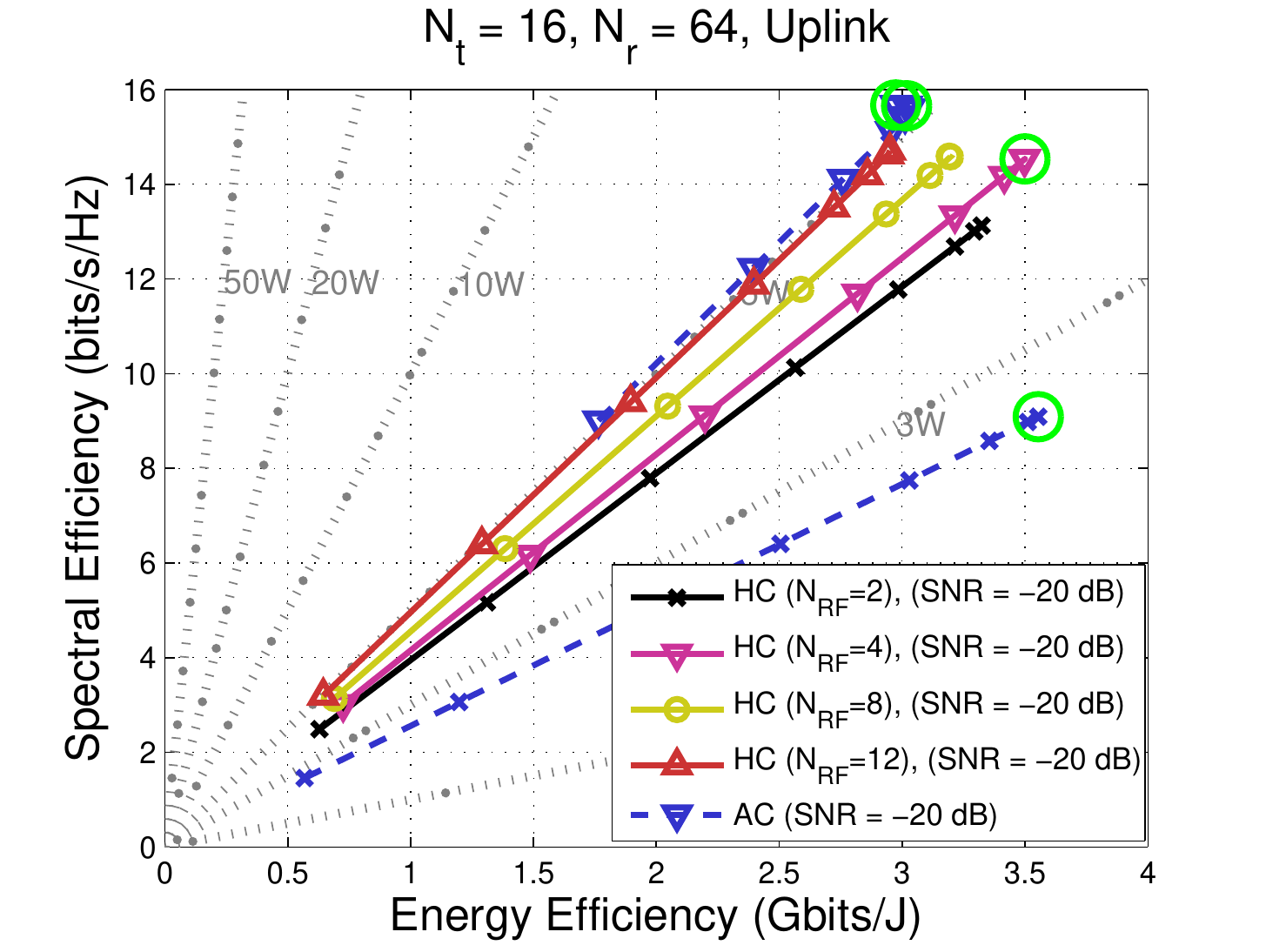}
    \label{fig:ULVarRFLSnrLPADC}
    }
  \caption{SE vs. EE comparison for AC, DC and HC schemes with a LPADC model and with $N_{RF} = 2,4,8,10,12$ for HC.}\vspace{-4mm}
    \label{fig:VarRFLPADC}
\end{figure*}

The HC receiver works much better in Uplink, as we can see comparing Figures \ref{fig:ULVarRFHSnr} and \ref{fig:ULVarRFLSnr}. Here we observe again that as $N_{RF}$ takes values $2, 4, 8 $ and $12$ the SE of HC grows at high SNR and becomes flat at low SNR. At high SNR, $N_{RF} = 8$ HC achieves the best EE and $N_{RF} = 12$ contributes a point with good EE-vs-SE trade-off, while DC provides other good points in the trade-off and result in the highest SE. However at low SNR $N_{RF}=4$ is the only HC configuration that outperforms DC in a trade-off point, and the best EE is again provided by AC. As in Downlink, note that the SNR dependence of the optimal $N_{RF}$ can be an issue for HC, and a wrong choice of $N_{RF}$ by a small difference makes the HC receiver worse than DC.

Finally, if we wish to add a constant device power consumption ruler we can superimpose it with diagonal dotted lines. In Downlink, only AC is viable under $1$ W, whereas above $1$ W DC outperforms the other schemes. In Uplink, only AC is viable under $3$ W, HC presents a better choice in the range $3-8$ W and DC outperforms all schemes above $8$ W. Surprisingly, DC is a better receiver for both smaller devices (UE with $1-3$ W) and large devices (macro cell BS with $10-50$ W), while HC is better for mid-range power devices (such as a pico cell BS with $3-10$W). In devices below 1 W AC seems to be the only viable option.

\subsection{Example 2: Low power ADC characteristics}

Note that all the observations made about Figure \ref{fig:VarRF} apply only to receivers fabricated with the component parameters in Table \ref{tab:devicepowers} and the HPADC in Table \ref{tab:adcparam}. A significant contribution of our analysis chart is its versatility to reproduce the analysis with changed parameters. In Figure \ref{fig:VarRFLPADC} we represent the same analysis performed with the LPADC model and $0$ W power consumption for phase shifters, as a sort of ``future expected values''. Since ADCs have lower power consumption, as $b$ is increased the lines progress to the top right and, differently from HPADC, EE always increases with $b$ in LPADC. Also, now the number of bits that maximizes SE for a receiver also maximizes EE.

In Fig. \ref{fig:DLVarRFHSnrLPADC} for high SNR Downlink we have a result in which DC completely outperforms all schemes and there is no trade-off. DC with $8$ bits is significantly above in terms of both EE and SE. For the Uplink in Fig. \ref{fig:ULVarRFHSnrLPADC} there is a slight trade-off in which points are very close. HC with $N_{RF}=8$ achieves slightly better EE with slightly worse SE, and DC with $b=7$ is chosen for some value of $\alpha$. 

To observe the impact of SNR in Downlink we compare Figures \ref{fig:DLVarRFHSnrLPADC} and \ref{fig:DLVarRFLSnrLPADC}. We observe again the saturation of HC SE as $N_{RF}$ grows from 2 to 10 at low SNR although it grows at high SNR. Moreover, DC remains the dominant receiver in most trade-off preferences except if one sets $\alpha=1$, where AC is chosen with a tiny gain in EE and a huge $50\%$ drop in SE. Cases such as AC here exemplify why a trade-off utility maximization interpretation can lead to new insights on the receiver selection problem.

Looking at the Uplink we compare the high and low SNR cases in Figures \ref{fig:ULVarRFHSnrLPADC} and \ref{fig:ULVarRFLSnrLPADC}, respectively. Here we observe again the saturation of HC with an increase of $N_{RF}$ in low SNR and the issue that the optimal $N_{RF}$ is SNR-dependent. In addition, mismatched $N_{RF}$'s are outperformed by DC. Furthermore, AC offers slightly better EE with a huge SE drop, HC contributes a couple of points in the trade-off, and DC contributes most of the points in the EE-vs-SE trade-off. 

Finally, with regard to power consumption, the same observations we made for HPADC apply here.

\section{Conclusions}
\label{sec:conc}
In this work, we studied and compared the spectral and energy efficiencies of Analog, Digital and Hybrid combining schemes. We considered receivers operating at mmWave and equipped with low resolution analog to digital converters (ADC). 
We developed a multi-objective optimization formulation, where the preference between SE and EE is weighted by a free parameter. This allows researchers/designers to identify the best receiver scheme depending on their needs.

We considered both Uplink and Downlink scenarios, and our results showed that in the Downlink, where the receiver is equipped with fewer antennas, DC outperforms other schemes, whereas in the Uplink there is a trade-off between HC and DC. Moreover, in low SNR scenarios AC achieves the highest EE, at the expense of severe drops in SE. 

We also argued that the EE vs SE comparison among different receiver designs is extremely parameter dependent, and results based on one chosen set of parameter values do not represent a complete picture due to the fact that ADCs and other component circuits are rapidly improving. To illustrate this, we produced comparison charts with low power and high power ADCs and showed that the SE vs EE trade-off results differ significantly. Therefore, ADCs and other component parameters directly affect the choice of the most appropriate combining scheme. To address the effect of this critical parameter dependency, we developed an easily reproduced analysis method in chart form and provided a web tool to complement this paper, where the readers may obtain their own SE vs EE trade-off charts by plugging in the set of parameter values for the hardware components they have available.
  

\input{EE_Cap_EW_UpIntro.bbl}

\end{document}

%% file: EE_Cap_EW_UpIntro.bbl

%% file: EE_Cap_EW_UpIntro.bbl
\begin{thebibliography}{10}
\providecommand{\url}[1]{#1}
\csname url@samestyle\endcsname
\providecommand{\newblock}{\relax}
\providecommand{\bibinfo}[2]{#2}
\providecommand{\BIBentrySTDinterwordspacing}{\spaceskip=0pt\relax}
\providecommand{\BIBentryALTinterwordstretchfactor}{4}
\providecommand{\BIBentryALTinterwordspacing}{\spaceskip=\fontdimen2\font plus
\BIBentryALTinterwordstretchfactor\fontdimen3\font minus
  \fontdimen4\font\relax}
\providecommand{\BIBforeignlanguage}[2]{{%
\expandafter\ifx\csname l@#1\endcsname\relax
\typeout{** WARNING: IEEEtran.bst: No hyphenation pattern has been}%
\typeout{** loaded for the language `#1'. Using the pattern for}%
\typeout{** the default language instead.}%
\else
\language=\csname l@#1\endcsname
\fi
#2}}
\providecommand{\BIBdecl}{\relax}
\BIBdecl

\bibitem{KhanFmmWave}
Z.~Pi and F.~Khan, ``An introduction to millimeter-wave mobile broadband
  systems,'' \emph{IEEE Communications Magazine}, vol.~49, no.~6, pp. 101--107,
  Jun. 2011.

\bibitem{5GWillWork}
T.~Rappaport, S.~Sun, R.~Mayzus, H.~Zhao, Y.~Azar, K.~Wang, G.~Wong, J.~Schulz,
  M.~Samimi, and F.~Gutierrez, ``Millimeter wave mobile communications for {5G}
  cellular: It will work!'' \emph{IEEE Access}, vol.~1, pp. 335--349, 2013.

\bibitem{mmWBF2014}
W.~Roh, J.~Y. Seol, J.~Park, B.~Lee, J.~Lee, Y.~Kim, J.~Cho, K.~Cheun, and
  F.~Aryanfar, ``Millimeter-wave beamforming as an enabling technology for {5G}
  cellular communications: theoretical feasibility and prototype results,''
  \emph{IEEE Communications Magazine}, vol.~52, no.~2, pp. 106--113, Feb. 2014.

\bibitem{AlkhateebMIMOSolMag}
A.~Alkhateeb, J.~Mo, N.~Gonzalez-Prelcic, and R.~W. Heath, ``{MIMO} precoding
  and combining solutions for millimeter-wave systems,'' \emph{IEEE
  Communications Magazine}, vol.~52, no.~12, pp. 122--131, Dec. 2014.

\bibitem{AyachHC}
O.~E. Ayach, S.~Rajagopal, S.~Abu-Surra, Z.~Pi, and R.~W. Heath, ``Spatially
  sparse precoding in millimeter wave {MIMO} systems,'' \emph{IEEE Transactions
  on Wireless Communications}, vol.~13, no.~3, pp. 1499--1513, Mar. 2014.

\bibitem{MoHeath1bit}
J.~Mo and R.~W. Heath, ``Capacity analysis of one-bit quantized {MIMO} systems
  with transmitter channel state information,'' \emph{IEEE Transactions on
  Signal Processing}, vol.~63, no.~20, pp. 5498--5512, Oct. 2015.

\bibitem{MassMIMO1bit}
S.~Jacobsson, G.~Durisi, M.~Coldrey, U.~Gustavsson, and C.~Studer, ``One-bit
  massive {MIMO}: Channel estimation and high-order modulations,'' in
  \emph{2015 IEEE International Conference on Communication Workshop (ICCW)},
  Jun. 2015.

\bibitem{confiwcmcNossekI06}
J.~A. Nossek and M.~T. Ivrlac, ``Capacity and coding for quantized {MIMO}
  systems.'' in \emph{Proceedings of the International Conference on Wireless
  Communications and Mobile Computing, (IWCMC 2006)}, 2006.

\bibitem{MurrayAGCQuant}
B.~M. Murray and I.~B. Collings, ``{AGC} and quantization effects in a
  zero-forcing {MIMO} wireless system,'' in \emph{IEEE 63rd Vehicular
  Technology Conference}, vol.~4, May 2006, pp. 1802--1806.

\bibitem{Madhow_1bitADC}
J.~Singh, O.~Dabeer, and U.~Madhow, ``On the limits of communication with
  low-precision analog-to-digital conversion at the receiver,'' \emph{IEEE
  Transactions on Communications}, vol.~57, no.~12, pp. 3629--3639, Dec. 2009.

\bibitem{OrhanER15PowerCons}
O.~Orhan, E.~Erkip, and S.~Rangan, ``Low power analog-to-digital conversion in
  millimeter wave systems: Impact of resolution and bandwidth on performance,''
  in \emph{Information Theory and Applications Workshop (ITA)}, Feb. 2015.

\bibitem{FanULRateLowADC15}
L.~Fan, S.~Jin, C.-K. Wen, and H.~Zhang, ``Uplink achievable rate for massive
  {MIMO} systems with low-resolution {ADC},'' \emph{IEEE Communications
  Letters}, vol.~19, no.~12, pp. 2186--2189, Dec. 2015.

\bibitem{ZhangSELowADC}
J.~Zhang, L.~Dai, S.~Sun, and Z.~Wang, ``On the spectral efficiency of massive
  {MIMO} systems with low-resolution {ADC}s,'' \emph{IEEE Communications
  Letters}, vol.~20, no.~5, pp. 842--845, May 2016.

\bibitem{Ahmed16_EE}
\BIBentryALTinterwordspacing
J.~Mo, A.~Alkhateeb, S.~Abu-Surra, and R.~W.~Heath~Jr, ``Hybrid architectures
  with few-bit {ADC} receivers: Achievable rates and energy-rate tradeoffs.''
  [Online]. Available: \url{https://arxiv.org/abs/1605.00668}
\BIBentrySTDinterwordspacing

\bibitem{MyPCCompEW16}
W.~b. Abbas and M.~Zorzi, ``Towards an appropriate receiver beamforming scheme
  for millimeter wave communication: A power consumption based comparison,'' in
  \emph{Proceedings of the 22nd European Wireless Conference, 2016}.

\bibitem{NYU_ADC65fJ}
\BIBentryALTinterwordspacing
B.~Nasri, S.~P. Sebastian, K.~You, R.~RanjithKumar, and D.~Shahrjerdi, ``A 700
  $\mu${W} 1 {GS}/s 4-bit folding-flash {ADC} in 65nm {CMOS} for wideband
  wireless communications.'' [Online]. Available:
  \url{https://arxiv.org/abs/1612.04855}
\BIBentrySTDinterwordspacing

\bibitem{FOM}
R.~H. Walden, ``Analog-to-digital converter survey and analysis,'' \emph{IEEE
  Journal on Selected Areas in Communications}, vol.~17, no.~4, pp. 539--550,
  Apr. 1999.

\bibitem{ADCs_97-15}
\BIBentryALTinterwordspacing
B.~Murmann, ``{ADC} performance survey 1997-2016.'' [Online]. Available:
  \url{http://web.stanford.edu/~murmann/adcsurvey.html}
\BIBentrySTDinterwordspacing

\bibitem{mmWaveADCwebviewer}
\BIBentryALTinterwordspacing
W.~b. Abbas, F.~Gomez-Cuba, and M.~Zorzi, ``mmwave receiver beamforming
  comparison tool.'' [Online]. Available:
  \url{http://enigma.det.uvigo.es/~fgomez/mmWaveADCwebviewer/}
\BIBentrySTDinterwordspacing

\bibitem{WaqasEConJour16}
\BIBentryALTinterwordspacing
W.~b. Abbas, F.~Gomez{-}Cuba, and M.~Zorzi, ``Millimeter wave receiver
  efficiency: {A} comprehensive comparison of beamforming schemes with low
  resolution {ADC}s,'' \emph{submitted to IEEE Transaction on Wireless
  Communication}, 2016. [Online]. Available:
  \url{http://arxiv.org/abs/1607.03725v1}
\BIBentrySTDinterwordspacing

\bibitem{Akdeniz_mmW_CM}
M.~R. Akdeniz, Y.~Liu, M.~K. Samimi, S.~Sun, S.~Rangan, T.~S. Rappaport, and
  E.~Erkip, ``Millimeter wave channel modeling and cellular capacity
  evaluation,'' \emph{IEEE Journal on Selected Areas in Communications},
  vol.~32, no.~6, pp. 1164--1179, Jun. 2014.

\bibitem{AQNM07}
A.~K. Fletcher, S.~Rangan, V.~K. Goyal, and K.~Ramchandran, ``Robust predictive
  quantization: Analysis and design via convex optimization,'' \emph{IEEE
  Journal of Selected Topics in Signal Processing}, vol.~1, no.~4, pp.
  618--632, Dec 2007.

\bibitem{ADC_b_B}
H.-S. Lee and C.~Sodini, ``Analog-to-digital converters: Digitizing the analog
  world,'' \emph{Proceedings of the IEEE}, vol.~96, no.~2, pp. 323--334, Feb.
  2008.

\bibitem{Rx_Pow_LNA_PS_C}
Y.~Yu, P.~Baltus, A.~de~Graauw, E.~van~der Heijden, C.~Vaucher, and A.~van
  Roermund, ``A 60 {GH}z phase shifter integrated with {LNA} and {PA} in 65 nm
  {CMOS} for phased array systems,'' \emph{IEEE Journal of Solid-State
  Circuits}, vol.~45, no.~9, pp. 1697--1709, Sep. 2010.

\bibitem{KongPhDPSPow}
\BIBentryALTinterwordspacing
L.~Kong, ``Energy-efficient 60 {GH}z phased-array design for multi-{G}b/s
  communication systems,'' \emph{Ph.D. dissertation, {EECS} Department,
  University of California, Berkeley}, Dec. 2014. [Online]. Available:
  \url{http://digitalassets.lib.berkeley.edu/techreports/ucb/text/EECS-2014-191.pdf}
\BIBentrySTDinterwordspacing

\bibitem{LPowPS0mW}
Y.-H. Lin and H.~Wang, ``A low phase and gain error passive phase shifter in 90
  nm {CMOS} for 60 {GH}z phase array system application,'' in \emph{2016 IEEE
  MTT-S International Microwave Symposium (IMS)}, May 2016, pp. 1--4.

\bibitem{Rx_Pow_60GHz}
M.~Kraemer, D.~Dragomirescu, and R.~Plana, ``Design of a very low-power,
  low-cost 60 {GH}z receiver front-end implemented in 65 nm {CMOS}
  technology,'' \emph{International Journal of Microwave and Wireless
  Technologies}, vol.~3, pp. 131--138, Apr. 2011.

\bibitem{RialandHeath}
R.~Mendez-Rial, C.~Rusu, N.~Gonzalez-Prelcic, A.~Alkhateeb, and R.~Heath,
  ``Hybrid {MIMO} architectures for millimeter wave communications: Phase
  shifters or switches?'' \emph{IEEE Access}, vol.~4, pp. 247--267, Jan. 2016.

\bibitem{TroppAPM05}
J.~A. Tropp, I.~S. Dhillon, R.~W. Heath, and T.~Strohmer, ``Designing
  structured tight frames via an alternating projection method,'' \emph{IEEE
  Trans. Inf. Theor.}, vol.~51, no.~1, pp. 188--209, Jan. 2005.

\end{thebibliography}
